# Magnetization dynamics and its scattering mechanism in thin CoFeB films with interfacial anisotropy

A. Okada[a,1], S. He[b,c,1], B. Gu[d], S. Kanai[a,e,f], A. Soumyanarayanan[b,c], S. T. Lim[c], M. Tran[c], M. Mori[d], S. Maekawa[d,g], F. Matsukura[a,e,f,h,2], H. Ohno[a,e,f,h,i], and C. Panagopoulos[b,2]

[a]Laboratory for Nanoelectronics and Spintronics, Research Institute of Electrical Communication, Tohoku University, 2-1-1 Katahira, Aoba-ku, Sendai 980-8577, Japan;
[b]Divison of Physics and Applied Physics, School of Physical and Mathematical Sciences, Nanyang Technological University, 637371, Singapore;
[c]Data Storage Institute, 5 Engineering Drive 1, 117608 Singapore;
[d]Advanced Science Research Center, Japan Atomic Energy Agency, Tokai 319-1195, Japan;
[e]Center for Spintronics Integrated Systems, Tohoku University, 2-1-1 Katahira, Aoba-ku, Sendai 980-8577, Japan;
[f]Center for Spintronics Research Network, Tohoku University, 2-1-1 Katahira, Aoba-ku, Sendai 980-8577, Japan;
[g]ERATO, Japan Science and Technoogy Agency, Sendai 980-8577, Japan;
[h]WPI-Advanced Institute for Materials Research, Tohoku University, 2-1-1 Katahira, Aoba-ku, Sendai 980-8577, Japan;
[i]Center for Innovative Integrated Electronic Systems, Tohoku University, 468-1 Aramaki Aza Aoba, Aoba-ku, Sendai 980-0845, Japan

**Significance**

Ferromagnetic resonance (FMR) is considered a standard tool in the study of magnetization dynamics, with an established analysis procedure. We show there is a missing piece of physics to consider in order to fully understand and precisely interpret FMR results. This physics manifests itself in the dynamics of magnetization and hence in FMR spectra of ferromagnets, where interfacial anisotropy is a fundamental term. Advances in ferromagnetic heterostructures enabled by developing cutting-edge technology now allow us to probe and reveal physics previously hidden in the bulk properties of ferromagnets.

**Studies of magnetization dynamics have incessantly facilitated the discovery of fundamentally novel physical phenomena, making steady headway in the development of magnetic and spintronics devices. The dynamics can be induced and detected electrically, offering new functionalities in advanced electronics at the nanoscale. However, its scattering mechanism is still disputed. Understanding the mechanism in thin films is especially important, because most spintronics devices are made from stacks of multilayers with nanometer thickness. The stacks are known to possess interfacial magnetic anisotropy, a central property for applications, whose influence on the dynamics remains unknown. Here, we investigate the impact of interfacial anisotropy by adopting CoFeB/MgO as a model system. Through systematic and complementary measurements of ferromagnetic resonance (FMR), on a series of thin films, we identify narrower FMR linewidths at higher temperatures. We explicitly rule out the temperature dependence of intrinsic damping as a possible cause, and it is also not expected from existing extrinsic scattering mechanisms for ferromagnets. We ascribe this observation to motional narrowing, an old concept so far neglected in the analyses of FMR spectra. The effect is confirmed to originate from interfacial anisotropy, impacting the practical technology of spin-based nanodevices up to room temperature.**

\body

The magnetization dynamics is determined by the combination of intrinsic and extrinsic effects. The intrinsic contribution is governed by the fundamental material parameter, damping constant $\alpha$ (1,2). The extrinsic counterpart is due to inhomogeneity and magnon excitations (3–5), hence, structure dependent and is enhanced in magnets with small thickness and/or small lateral dimensions (4–7). The contribution from each effect is usually separated by the analysis of the linewidths of FMR spectra, in which the linewidth enhancement is caused by the distributions of magnitudes and directions of the effective magnetic anisotropy, and magnon excitations. In this work, we study the temperature and CoFeB thickness dependences of the FMR linewidths in CoFeB/MgO thin films, one of the most promising material systems for high-performance spintronics devices at the nanoscale. The system possesses sizable interfacial perpendicular magnetic anisotropy (8), whose effect on the FMR linewidth is the focus of this study.

**Samples and measurement setups**

We prepared thin $Co_{0.4}Fe_{0.4}B_{0.2}$ layers with thickness $t$ ranging from 1.4 to 3.7 nm, sandwiched between 3 nm thick MgO layers using magnetron sputtering. The thicknesses of the layers were calibrated by transmission electron microscopy. All the samples studied in this work possess in-plane easiness for magnetization. FMR spectra were measured both by a vector-network-analyzer (VNA-FMR) using a coplanar waveguide (CPW), and a conventional method using a $TE_{011}$ microwave cavity (cavity-FMR). The former technique enables us to measure the rf frequency $f$ dependence of the spectra up to 26 GHz by applying an external magnetic field $H$ either parallel (magnetic field angle $\theta_H = 90°$) or perpendicular ($\theta_H = 0°$) to the sample plane, and the spectra are obtained as the transmission coefficient $S_{21}$ (9). The latter technique measures the spectra at a fixed $f$ of 9 GHz under $H$ at various $\theta_H$, and the spectra are obtained as the derivative of the microwave absorption with respect to $H$ (10). The temperature dependence of spontaneous magnetization $M_S$ was measured by a superconducting quantum interference device (SQUID) magnetometer.

## VNA-FMR

Fig. 1A shows typical VNA-FMR at selected values of $f$ for CoFeB with $t = 1.5$ nm at temperature $T = 300$ K and $\theta_H = 90°$. We determine the resonance field $H_R$ and linewidth (full width at half maximum) $\Delta H$ from the fitting of the modified Lorentz function (solid lines in Fig. 1A) to the VNA-FMR spectra (9). Fig. 1B shows the rf frequency dependence of $H_R$ at $\theta_H = 0°$ and $90°$, which we use to determine the effective perpendicular anisotropy fields $H_K^{eff}$ from the resonance condition; $f = \mu_0\gamma(H_R+H_K^{eff})/(2\pi)$ for $\theta_H = 0°$ and $f = \mu_0\gamma[H_R(H_R-H_K^{eff})]^{1/2}/(2\pi)$ for $\theta_H = 90°$. Here, $\mu_0$ is the permeability in free space, and $\gamma$ the gyromagnetic ratio. As shown in Fig. 1C, $H_K^{eff}$ increases monotonically with decreasing $t$, indicative of interfacial perpendicular anisotropy at the CoFeB/MgO interface (8).

Figs. 1D-1G show the frequency dependence of $\Delta H$ as a function of $t$ and $T$. As shown in Fig. 1D, when $H$ is applied perpendicular to the film ($\theta_H = 0°$), $\Delta H$ obeys linear dependence, which is expressed as $\Delta H = \Delta H_{in}+\Delta H_{inhom} = (2h\alpha/g\mu_0\mu_B)f+\Delta H_{inhom}$, where $h$ is the Planck constant and $\mu_B$ is the Bohr magneton (9,11). Here, $\Delta H_{in}$ is related to the intrinsic linewidth governed by $\alpha$ and $\Delta H_{inhom}$ is the extrinsic contribution due to inhomogeneity, such as the distribution of magnetic anisotropy. The value of $\alpha$ determines the slope in Fig. 1D and $\Delta H_{inhom}$ corresponds to the intercept on the vertical axis. Fig. 1D shows that $\alpha$ is nearly independent of $t$ ($\alpha \sim 0.004$) and $\Delta H_{inhom}$ increases with decreasing $t$. As seen from Fig. 1E for CoFeB with $t = 1.5$ nm, $\alpha$ is also nearly independent of $T$, whereas $\Delta H_{inhom}$ increases with decreasing $T$. It is known that $\alpha$ depends on $T$ through the change in resistivity with $T$ (12–14). The nearly temperature-independent $\alpha$ observed here may be due to the small temperature dependence of the resistivity of CoFeB (15). When $H$ is in-plane ($\theta_H = 90°$) (Fig. 1F), we observe a non-linearity, which is enhanced with decreasing $t$. The non-linearity, so far, is believed to be due to the contribution of two-magnon scattering (TMS) to the linewidth $\Delta H_{TMS}$; TMS is known to be activated when the magnetization angle $\theta_M$ from the sample normal is greater than $45°$ (4–6). The non-linear frequency dependence of $\Delta H$ in Fig. 1F, can be described in terms of TMS (dashed lines), assuming $\Delta H = \Delta H_{in}+\Delta H_{inhom}+\Delta H_{TMS}$. As depicted in Fig. 1G, the non-linearity for CoFeB with $t = 1.5$ nm is enhanced strongly with decreasing $T$ (nearly twice at 80 K compared to 300 K). This strong temperature dependence cannot be attributed to TMS, because the change in $\Delta H_{TMS}$ with $T$ from 300 to 4 K is calculated to be ~10% at most, using the measured $T$ dependence of $M_S$ and $H_K^{eff}$ (4). Hence, it is imperative to consider an alternative mechanism for the non-linearity, possibly related to the CoFeB/MgO-interface effect; notably, the non-linearity is absent in thicker CoFeB with $t = 3.7$ nm (Figs. 1F and 1G).

**Cavity-FMR**

Fig. 2A shows typical cavity-FMR spectra at $T = 300$ K for CoFeB with $t = 1.5$ nm. Fitting the derivative of the Lorentz function to the spectra gives $H_R$ and $\Delta H$ (16). We fit the resonance condition to the magnetic-field angle dependence of $H_R$ shown in Fig. 2B to obtain the magnitude of the effective first-order and second-order perpendicular anisotropy fields, $H_{K1}^{eff}$ and $H_{K2}$, by following the procedure in Ref. 10. Fig. 2C shows the CoFeB thickness dependence of $H_{K1}^{eff}$ and $H_{K2}$ along with that of $H_K^{eff}$ in Fig. 1C obtained from VNA-FMR. $H_{K2}$ is nearly independent of $t$ and its strength is a few tens of mT at most. $H_{K1}^{eff}$ increases with decreasing $t$, in agreement with $H_K^{eff}$ obtained from VNA-FMR, confirming again the presence of perpendicular interfacial anisotropy at the CoFeB/MgO interface (8). In Fig. 2C, we plot also the values of $H_K^{eff}$ obtained from magnetization measurements (8), which show good correspondence with those obtained from FMR measurements. Fig. 2D shows the magnetic-field angle dependence of $\Delta H$ as a function of $t$. For $\theta_M > 45°$, where TMS is expected to be activated, $\Delta H$ is larger for CoFeB with smaller $t$. As shown in Fig. 2E, however, a nearly twice larger $\Delta H$ at lower $T$ cannot be explained by the TMS contribution to $\Delta H$. The results obtained from cavity-FMR are consistent with those from VNA-FMR, indicating that the unexpected $T$ dependence of $\Delta H$ is not an artifact.

## Discussion

Because the linewidth enhancement for $\theta_H = 90°$ is larger for thinner CoFeB (Figs. 1F and 2D), it is natural to consider that its mechanism is related to an interfacial effect in CoFeB/MgO. The most pronounced interfacial effect is the presence of interfacial perpendicular anisotropy as seen in Figs. 1C and 2C. Fig. 2F shows $\log(K(T)/K(4\text{ K}))$ versus $\log(M_S(T)/M_S(4\text{ K}))$ (Callen-Callen plot) (17), where $M_S(T)$ is the measured spontaneous magnetization and $K(T)$ the perpendicular magnetic anisotropy energy density determined from $H_{K1}^{\text{eff}}$ and $H_{K2}$ using cavity-FMR (10); $K = M_S(H_{K1}^{\text{eff}} - H_{K2}/2)/2 + M_S^2/(2\mu_0)$. The linear behavior in Fig. 2F gives a slope $m$ of 2.16 (the exponent $m$ in the Callen-Callen law of $K(T)/K(0) = (M_S(T)/M_S(0))^m$), consistent with the relationship between interfacial anisotropy energy density and $M_S$ reported for CoFeB/MgO systems (18–20). Hence, the temperature dependence of anisotropy in our CoFeB films is also governed by the interfacial anisotropy. Therefore, the linewidth broadening with decreasing $t$ and $T$ is expected to be due to random thermal fluctuation $\delta H_i$ in the anisotropy field at interfacial site $i$ of a magnetic atom. Indeed, phonons can induce thermal fluctuations of interfacial anisotropy through vibrations of interfacial atoms. Because $\delta H_i$ is along the film normal (the in-plane component is expected to cancel out), it gives rise to a broader linewidth at larger $\theta_M$, which is similar to the angular dependence of the TMS contribution. The $T$ dependence of linewidths is also explained by the presence of $\delta H_i$. The value of $\delta H_i$ fluctuates with time due to the thermal fluctuation of phonons, the frequency of which increases with increasing $T$, resulting in motional narrowing by averaging the randomness of $\delta H_i$, albeit an increase in its amplitude with increasing $T$ (21).

To formulate the motional narrowing, we apply the Holstein-Primakov, Fourier, and Bogolyubov transformations to the spin-deviation Hamiltonian (22), and obtain $H' = -\Sigma_i \delta \mathbf{H}_i \cdot \mathbf{S}_{iz} \approx (S/2)^{1/2} \sin\theta_M \Sigma_\mathbf{k} \delta H_{-\mathbf{k}} (u_\mathbf{k} + v_\mathbf{k})(\alpha_\mathbf{k} + \alpha^+_{-\mathbf{k}})$ ($z$ direction along film normal) (see Materials and Methods). $H'$ describes the coupling between local spin $\mathbf{S}_i$ and fluctuating field $\delta H_i$, which contributes to spin relaxation and thus FMR linewidth. Here, $\mathbf{k}$ is the wavevector of magnon, $\alpha_\mathbf{k}$, $\alpha^+_{-\mathbf{k}}$, $u_\mathbf{k}$, and $v_\mathbf{k}$ are the annihilation and creation operators for magnon and their coefficients after the Bogolyubov transformation, and $S$ is the magnitude of spin. Adopting the Redfield theory to obtain the relationship between spin relaxation time and $\delta H_i$ (21), the linewidth $\Delta H_{MN}$ due to $\delta H_i$ at $k = 0$ (Kittel mode) is expressed as, $\Delta H_{MN} \approx (S/2)(\delta H_{k=0})^2 \sin^2\theta_M (H_1/H_2)^{1/2}$ with $H_1 = H_R\cos(\theta_H-\theta_M)+H_{K1}^{eff}\cos^2\theta_M-H_{K2}\cos^4\theta_M$, $H_2 = H_R\cos(\theta_H-\theta_M)+H_{K1}^{eff}\cos 2\theta_M-(H_{K2}/2)(\cos 2\theta_M+\cos 4\theta_M)$, and $(\delta H_{k=0})^2 = \int d\tau \delta H_{k=0}(t_0)\delta H_{k=0}(t_0+\tau)e^{-2i\pi f\tau}$. Here, $t_0$ is the arbitrary time, and $\tau$ is the elapsed time from $t_0$. Writing the correlation function $\delta H_{k=0}(t_0)\delta H_{k=0}(t_0+\tau) = (\delta H)^2 e^{-|\tau|/\tau_0}$, where $\tau_0$ the relaxation time of the random field $\delta H_{k=0}$, we obtain,

$$\Delta H_{MN} \approx (S/2)(\delta H)^2 \tau_0 \sin^2\theta_M (H_1/H_2)^{1/2} = \Gamma \sin^2\theta_M (H_1/H_2)^{1/2}, \qquad [1]$$

for $2\pi f \tau_0 \ll 1$.

We fit

$$\Delta H = \Delta H_{in}(\alpha, H_{K1}^{eff}, H_{K2}) + \Delta H_{inhom}(\Delta H_{K1}^{eff}, \Delta H_{K2})$$
$$+ \Delta H_{TMS}(M_S, H_{K1}^{eff}, H_{K2}, A_S, A, N) + \Delta H_{MN}(\Gamma, H_{K1}^{eff}, H_{K2}), \qquad [2]$$

to the magnetic-field angle dependence of the linewidths as a function of $T$ in Fig. 2E (3, 10). Each contribution in the right-hand side of Eq. 2 is determined by the parameters in parentheses. The values of $M_S$, $\alpha$, $H_{K1}^{eff}$, and $H_{K2}$ and their temperature dependence are determined experimentally; $M_S$ from magnetization measurements, $\alpha$ obtained from VNA-FMR measurements at $\theta_H = 0°$ (Fig. 3A), and $H_{K1}^{eff}$ and $H_{K2}$ from cavity-FMR measurements. We calculate $\Delta H_{in}$ from $\Delta H_{in} = \alpha(H_1+H_2)|dH_R/d[(H_1H_2)^{1/2}]|$ (3). The contribution from $\Delta H_{inhom}$ is expressed as $\Delta H_{inhom} = |dH_R/dH_{K1}^{eff}|\Delta H_{K1}^{eff}+|dH_R/dH_{K2}|\Delta H_{K2}$, and $\Delta H_{K1}^{eff}$ and $\Delta H_{K2}$ are adopted as fitting parameters (10). To describe the contribution from TMS, we adopt the expression in Ref. 4. The TMS contribution at 300 K is determined from the best fit of Eq. 2 to the experimental result at 300 K with two defect-related fitting parameters, $A$ and $N$, which reflect the size and density as well as the aspect ratio of the defects, respectively. The $T$ dependence of TMS is calculated using the $T$ dependence of $M_S$ and $H_K^{eff}$, assuming the exchange stiffness constant $A_S(T) \propto [M_S(T)]^2$ and $T$ independent $A$ and $N$ (20). As described before, the calculated TMS contribution changes by ~10% at most with decreasing $T$ from 300 to 4 K. For the $\Delta H_{MN}$ contribution, we adopt $\Gamma$ as an adjustable parameter. The fit agrees with the experimental results as shown by the solid lines in Fig. 2*E*. Fig. 2*G* depicts the calculated linewidths as functions of $f$ and $T$ using parameters obtained from the analyses, reproducing the results in Fig. 1*G*. We note that the difference in the detected areas between VNA-FMR (~0.1 mm$^2$) and cavity-FMR (~20 mm$^2$) may result in the small difference observed in the magnitude of $\Delta H$ (compare Fig. 1*G* and Fig. 2*G*), due to inhomogeneity. The $T$ dependence of $\Gamma$ is shown in Fig. 3*B*. Although the observed functional form is unknown, it may be due to several contributions, such as the $T$ dependence of the magnitude of $\delta H_i$ and the magnon and phonon lifetimes (23). $\delta H_i$ is expected to be determined by the $T$ dependence of the thermal lattice expansion coefficient and the resultant lattice mismatch between CoFeB and MgO (24), whereas the magnon lifetime may be due to the exchange/stiffness constants at the interface (25,26).

Furthermore, we fit Eq. **2** to the angle dependence of the linewidth for CoFeB with different $t$. Here, we treat $\alpha$, $\Delta H_{K1}^{eff}$, $\Delta H_{K2}$, TMS parameters, and $\Gamma$ as adjustable parameters. The CoFeB thickness dependence of $\alpha$ is shown in Fig. 3*C*. The magnitude of $\alpha$ is ~0.004 independent of $t$ (closed circles), and consistent with the values obtained from VNA-FMR spectra at $\theta_H = 0°$ (open circles). If we neglect $\Delta H_{MN}$ in the analyses of cavity-FMR linewidths, $\alpha$ increases with decreasing $t$ (squares). It is therefore, essential to include $\Delta H_{MN}$ in FMR analyses in order to obtain accurate values of $\alpha$ in thin CoFeB/MgO when $\theta_H \neq 0°$. Crucially, for the present CoFeB/MgO systems, we design the stacks to suppress the spin pumping effect by sandwiching CoFeB by two MgO layers (27,28). If we replace one or two sides of the adjacent MgO with Ta, $\alpha$ increases rapidly with decreasing $t$ due to spin pumping (triangles and diamonds). The $t$ dependence of $\Gamma$ in Fig. 3*D* attests to its interfacial origin.

From systematic FMR studies, we have shown that interfacial anisotropy in thin film CoFeB/MgO has a strong effect on the spectral linewidth. This effect is explained in terms of motional narrowing, which is commonly neglected in the analysis of FMR spectra. The present investigation demonstrates that great care must be taken in the study of FMR in magnetic architectures with interfacial anisotropy, a fundamental property for spin-based device applications. In addition, the result is expected to bring a new concept to spintronics devices utilizing phonon-magnon coupling through the interfacial anisotropy (29,30).

**Materials and Methods**

**Sample preparation.** Films were deposited on a thermally oxidized Si substrate by ultrahigh-vacuum magnetron sputtering. The stack structure, from substrate side, is Ta(5)/ CuN(30)/ TaN(20)/ Ta(5)/ [CoFeB(0.6)/ Ru(3)]$_2$/ Ta(2)/ CoFeB(0.6)/ MgO(3)/ CoFeB($t$ = 1~2.6)/ MgO(3)/ Ta(5)/ Ru(5)/ CuN(15), where numbers in parentheses are thickness in nm. The three 0.6-nm-thick CoFeB layers were inserted to improve the quality of the films above them (31,32), and are expected to exhibit superparamagnetic behavior (33). We confirmed they do not affect the FMR spectra using a reference measurement on a stack in their absence. Because they are expected to exhibit very different anisotropy, due to different interfaces and thickness (compared to thick CoFeB, which is of interest here) their resonances are not detected in the temperature, frequency and field range investigated. The FMR active layer is CoFeB sandwiched between two MgO layers. The CoFeB thickness $t$ is varied from 1 to 2.6 nm over 8" wafer using wedged-film deposition. The actual thicknesses of CoFeB are calibrated from cross-sectional images obtained using a transmission electron microscope. We prepared also a reference sample with thick CoFeB namely, $t$ = 3.7 nm.

**VNA-FMR.** Most of the measurements were performed using a home-built instrument with applied magnetic fields up to 0.55 T (up to 1.4 T at 300 K) and frequencies up to 26 GHz. A separate home-built FMR dipper probe was used to measure the sample with $t$ = 1.5 nm under a perpendicular magnetic field at temperatures down to 4.2 K. FMR spectra were acquired by sweeping the external magnetic field (9).

**Cavity-FMR.** The sample was placed in a TE$_{011}$ microwave cavity, where microwave frequency $f$ = 9 GHz was introduced. We measured the external magnetic-field $H$ dependence of the FMR spectrum (derivative microwave absorption spectrum) by superimposing an a.c. magnetic field (1 mT and 100 kHz) for lock-in detection. The sample temperature was controlled from 4 to 300 K using a liquid He flow cryostat (10, 16).

**Motional Narrowing.** The magnetic energy in the CoFeB film is described by a Heisenberg Hamiltonian $H_H$ with anisotropy term $D$ and external magnetic field $H_0$,

$$H_H = -J\sum_{\langle i,j\rangle}\mathbf{S}_i\cdot\mathbf{S}_j - D\sum_i S_{iz}^2 - \mathbf{H}_0\cdot\sum_i \mathbf{S}_i. \quad [3]$$

To study the linewidth in FMR due to the fluctuating field $\delta H_i$ at interfaces, we consider the perturbation Hamiltonian,

$$H' = -\sum_i \delta H_i S_{iz}, \quad [4]$$

here we consider $\delta H_i$ along the normal of the interfaces ($z$ direction) because its in-plane components are expected to cancel out. We study the case for tilted magnetization direction to in-plane direction along $x$, and define the direction of magnetization **M** as $\zeta$ axis. By rotating by $\theta_M$ about $y$ axis, we convert $x$ and $z$ axes of Cartesian coordinate system to $\xi$ and $\zeta$ axes. In the $\xi$–$\zeta$ plane, the spin operator in the $z$ direction is given by,

$$S_{iz} = S_{i\zeta}\cos\theta_M - S_{i\xi}\sin\theta_M, \quad [5]$$

and Eq. 4 is rewritten as,

$$H' = -\sum_i \delta H_i(S_{i\zeta}\cos\theta_M - S_{i\xi}\sin\theta_M), \quad [6]$$

Assuming a constant longitudinal spin component (along $\zeta$ direction) $S_{i\zeta}$, only the transverse spin component (along $\xi$ direction) $S_{i\xi}$ contributes to FMR. Hence, a Hamiltonian contributing to FMR is expressed as,

$$H'_\xi = \sin\theta_M\sum_i \delta H_i S_{i\xi}. \quad [7]$$

By the Holstein-Primakoff transformation with the creation and annihilation operators $a^+_i$ and $a_i$, and the magnitude of spin $S$, the transverse component is written as (22,34),

$$S_{i\xi} = \sqrt{\frac{S}{2}}(a_i + a_i^+),  \quad [8]$$

and Eq. 7 becomes,

$$H'_\xi = \sqrt{\frac{S}{2}} \sin\theta_M \sum_i \delta H_i (a_i + a_i^+), \quad [9]$$

The Fourier transformation with wavevector **k** and number of interfacial sites $N$ gives,

$$a_i = \frac{1}{\sqrt{N}} \sum_{\mathbf{k}} e^{i\mathbf{k}\cdot\mathbf{r}_i} a_{\mathbf{k}}, \quad [10]$$

$$\delta H_i = \frac{1}{\sqrt{N}} \sum_{\mathbf{k}} e^{i\mathbf{k}\cdot\mathbf{r}_i} \delta H_{\mathbf{k}}. \quad [11]$$

Then, Eq. 5 is expressed as,

$$H'_\xi = \sqrt{\frac{S}{2}} \sin\theta_M \sum_{\mathbf{k}} \delta H_{-\mathbf{k}} (a_{\mathbf{k}} + a_{-\mathbf{k}}^+). \quad [12]$$

Eq. 3 is diagonalized by the Bogolyubov transformation (22,35),

$$a_{\mathbf{k}} = u_{\mathbf{k}} \alpha_{\mathbf{k}} + v_{\mathbf{k}} \alpha^+_{-\mathbf{k}}, \quad [13]$$

$$a_{-\mathbf{k}} = v_{\mathbf{k}} \alpha^+_{\mathbf{k}} + u_{\mathbf{k}} \alpha_{-\mathbf{k}}, \quad [14]$$

and Eq.12 transforms to,

$$H'_\xi = \sqrt{\frac{S}{2}} \sin\theta_M \sum_{\mathbf{k}} \delta H_{-\mathbf{k}} (u_{\mathbf{k}} + v_{\mathbf{k}})(\alpha_{\mathbf{k}} + \alpha^+_{-\mathbf{k}}). \quad [15]$$

Because we are interested in the FMR mode, we consider the **k** = 0 mode in Eq. 15,

$$H'_\xi(\mathbf{k}=0) = \sqrt{\frac{S}{2}} \sin\theta_M \delta H_{\mathbf{k}=0} (u_0 + v_0)(\alpha_{\mathbf{k}=0} + \alpha^+_{\mathbf{k}=0}). \quad [16]$$

Adopting standard process to obtain the magnon dispersion, the coefficients in the Bogolyubov transformation in Eq. 16 are obtained as,

$$(u_0 + v_0)^2 = \sqrt{\frac{H_1}{H_2}}, \qquad [17]$$

with

$$H_1 \equiv H_0 \cos(\theta_H - \theta_M) + 2DS\cos^2\theta_M, \qquad [18]$$

$$H_2 \equiv H_0 \cos(\theta_H - \theta_M) + 2DS\cos 2\theta_M. \qquad [19]$$

According to Redfield theory (21,36), the linewidth of FMR (**k** = 0) mode is proportional to the spectral density of fluctuating fields $\delta H_{\mathbf{k}=0}$,

$$\Delta H_{MN} \approx \frac{S}{2}\sin^2\theta_M (u_0+v_0)^2 \int_0^\infty \delta H_{\mathbf{k}=0}(t)\delta H_{\mathbf{k}=0}(t+\tau) e^{-i\omega\tau} d\tau, \qquad [20]$$

where the time correlation function of fluctuating field $\delta H_{\mathbf{k}=0}$ at interfaces is given by (21),

$$\delta H_{\mathbf{k}=0}(t)\delta H_{\mathbf{k}=0}(t+\tau) = (\delta H)^2 e^{-|\tau|/\tau_0}. \qquad [21]$$

Because only *z* component of $\delta H$ is relevant, we assume the single relaxation lifetime $\tau_0$ of the fluctuating fields at interfaces. $\tau_0$ is of the same order of the inverse of the phonon frequency - much larger than the resonance frequency $\omega$, and thus $\omega\tau_0 \ll 1$. The integration over time $\tau$ is calculated as,

$$\int_0^\infty e^{-|\tau|/\tau_0} e^{-i\omega\tau} d\tau = \frac{\tau_0}{1+\omega^2\tau_0^2} \xrightarrow{\omega\tau_0 \ll 1} \tau_0. \qquad [22]$$

Finally, we obtain the following linewidth for the FMR mode due to the fluctuating fields at interfaces,

$$\Delta H_{MN} \approx \frac{S}{2}(\delta H)^2 \tau_0 \sin^2\theta_M (H_1/H_2)^{1/2}, \qquad [23]$$

which is Eq. **1** in the main text.

**ACKNOWLEDGMENTS.** The work at TU was supported in part by Grants-in-Aid for Scientific Research from MEXT (No. 26103002) and from JSPS (Nos. 16H06081 and 16J05455), R&D project for ICT Key Technology of MEXT, ImPACT program of CSTI, JSPS Core-to-Core Program, as well as the Cooperative Research Projects of RIEC. The work in Singapore was supported by the Ministry of Education (MOU, AcRF Tier 2 Grant (MOE2014-T2-1-050), The A*STAR Pharos Fund (1527400026), and the National Research Foundation (NRF), NRF-Investigatorship (NRFNRFI2015-04). The work at JAEA was supported in part by Grants-in-Aid for Scientific Research from JSPS (Nos. 26247063 and 25287094) and from MEXT (26103006).

**Figure Captions**

**Fig. 1.** Vector-network-analyzer ferromagnetic resonance (VNA-FMR). (*A*) Typical spectra (real part of transmission coefficient Re($S_{21}$) of coplanar waveguide) of CoFeB with thickness $t$ = 1.5 nm as a function of rf frequency $f$ obtained at magnetic–field angle $\theta_H$ = 90º and temperature $T$ = 300 K. Solid lines are fits by the modified Lorentzian. (*B*) rf frequency $f$ dependence of resonance field $H_R$ obtained at $\theta_H$ = 0º and 90º. (*C*) Thickness $t$ dependence of effective perpendicular anisotropy fields $H_K^{eff}$. rf frequency $f$ dependence of the FMR linewidths $\Delta H$ obtained at $\theta_H$ = 0º as a function of (*D*) $t$ at $T$ = 300 K and as a function of (*E*) $T$ for CoFeB with $t$ = 1.5 nm. Solid lines in Figs. 1*D* and *E* are linear fits. (*F*), (*G*) Same as Fig. 1*D*, *E* but at $\theta_H$ = 90º. Dashed lines in Figs. 1*F* and *G* are non-linear fits based on two-magnon scattering.

**Fig. 2.** Cavity ferromagnetic resonance (Cavity-FMR). (*A*) Typical spectra of CoFeB with thickness $t$ of 1.5 nm as a function of magnetic–field angle $\theta_H$ at temperature $T$ = 300 K. Solid lines are fits by the derivative of Lorentzian. (*B*) Angle $\theta_H$ dependence of resonance field $H_R$ as a function of $t$. Solid lines are fitted lines by using the resonance condition. (*C*) Thickness $t$ dependence of effective first-order perpendicular anisotropy field $H_{K1}^{eff}$ and second-order anisotropy field $H_{K2}$ along with the results in Fig. 1*C* and from magnetization measurements. Angle $\theta_H$ dependence of the FMR linewidth $\Delta H$ obtained as a function of (*D*) $t$ at $T$ = 300 K and as a function of (*E*) $T$ for CoFeB with $t$ = 1.5 nm. Solid lines in Figs. 2*D* and *E* are fitted lines. (*F*) Double-logarithm plot of normalized magnetic anisotropy energy density $K(T)/K(4K)$ versus normalized spontaneous magnetization $M_S(T)/M_S(4\ K)$. Solid line is a linear fit. (*G*) Calculated $\Delta H$ as a function of rf frequency $f$ at different temperatures by using the parameters obtained from the analyses of cavity-FMR spectra.

**Fig. 3.** Temperature $T$ and CoFeB thickness $t$ dependences of damping constant $\alpha$ and linewidths $\Gamma$ relating to motional narrowing. Temperature $T$ dependence of (*A*) $\alpha$ and (*B*) $\Gamma$ for CoFeB with $t$ = 1.5 nm. Thickness $t$ dependence of (*C*) $\alpha$ and (*D*) $\Gamma$. Filled symbols are determined from cavity-FMR and open symbols from VNA-FMR.

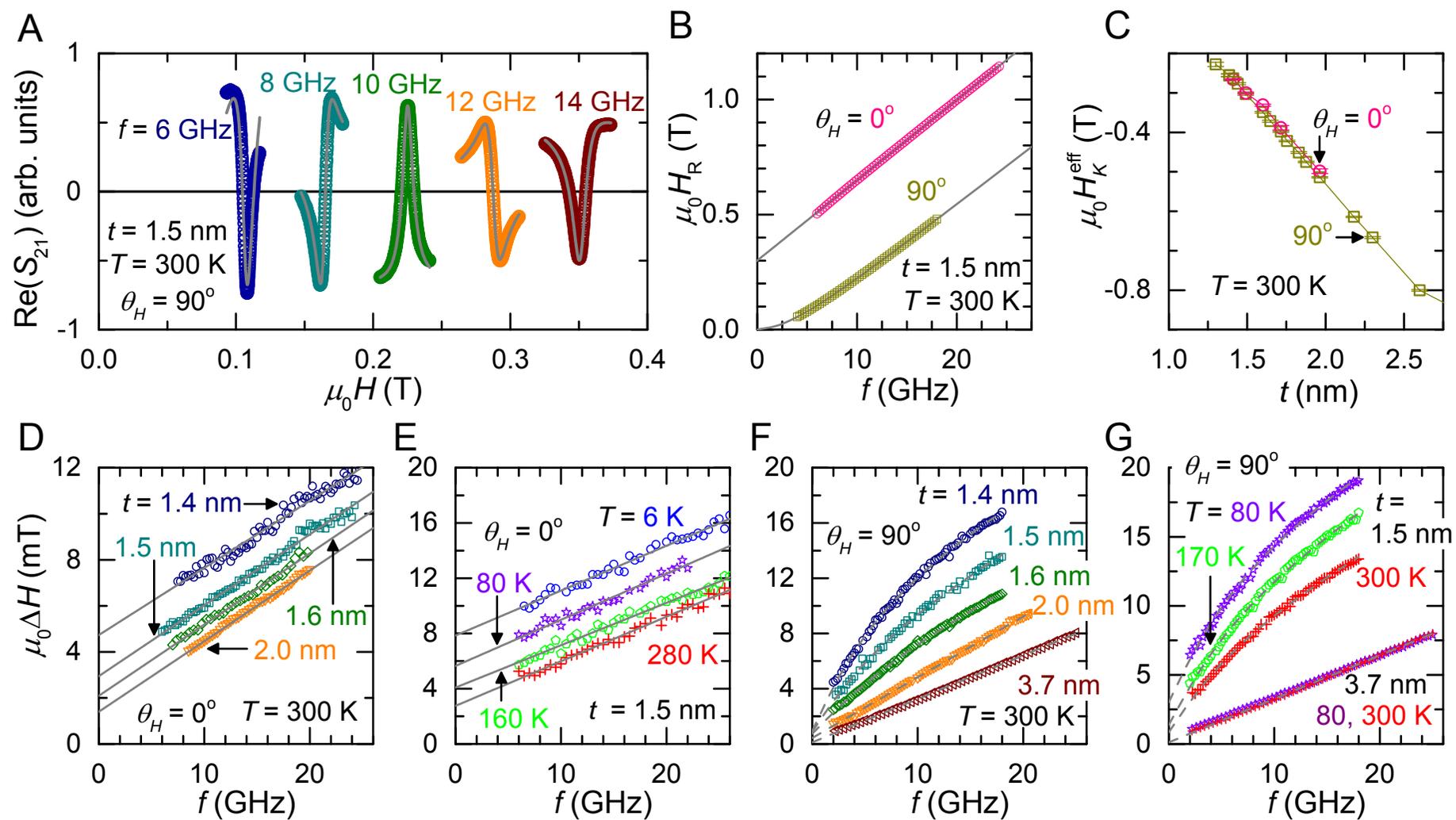

Fig. 1 Okada *et al.*

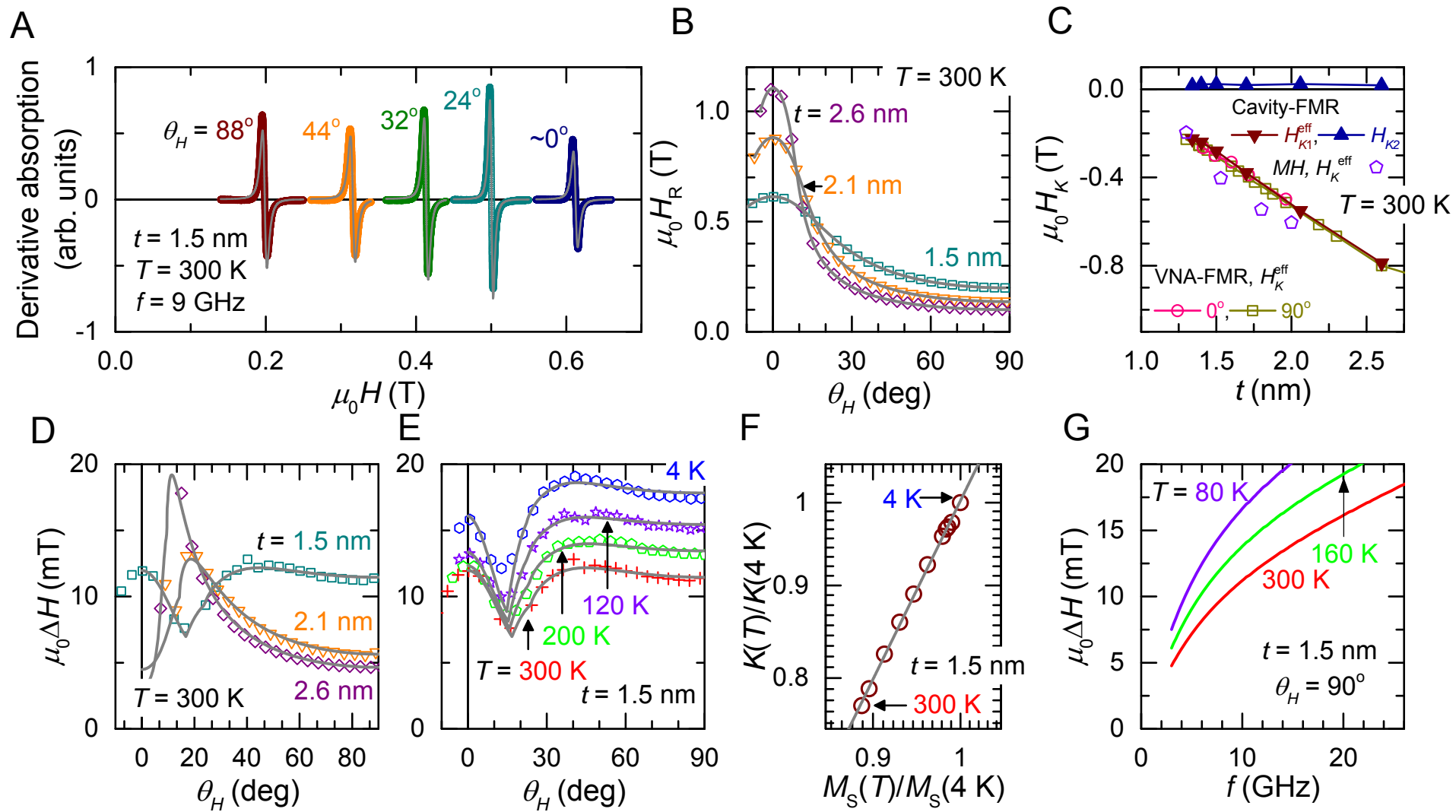

Fig. 2 Okada *et al.*

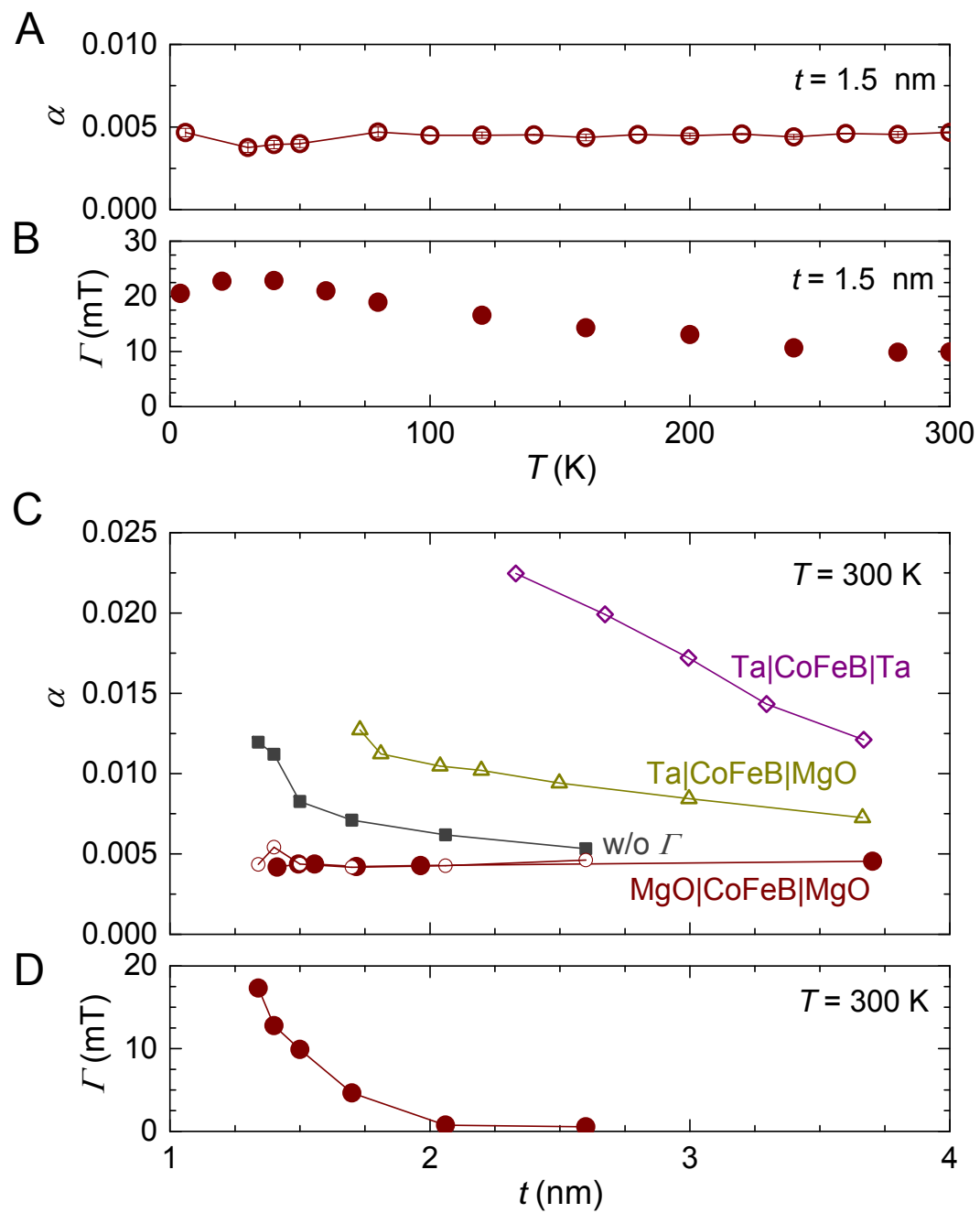

Fig. 3 Okada *et al.*